\documentclass[aps,prb,twocolumn,footinbib,showpacs]{revtex4}         
\usepackage{graphicx}
\usepackage{amsmath}

\begin{document}

\font\type=cmtt12
\def\undertext#1{$\underline{\smash{\hbox{#1}}}$}
\def\summary #1{\hbox to 16cm{\hskip 2cm\vtop{\hbox{\vbox{\hsize 12cm
                         \vskip 10pt
                         \noindent #1
                         \vskip 10pt
                           }}}}
           }
\def\ie{{\it ie}} \def\ea{{\it et al.}} \def\eg{{\it e.g.}} 
\def\cline{\centerline} \def\lline{\leftline} \def\rline{\rightline}
\def\dhalf{{{\scriptstyle 1}\over{\scriptstyle 2}}}
\def\dsixth{{{\scriptstyle 1}\over{\scriptstyle 6}}}
\def\half{{{\scriptscriptstyle 1}\over{\scriptscriptstyle 2}}}
\def\sixth{{{\scriptscriptstyle 1}\over{\scriptscriptstyle 6}}}
\def\quarter{{{\scriptscriptstyle 1}\over{\scriptscriptstyle 4}}}
\def\third{{{\scriptscriptstyle 1}\over{\scriptscriptstyle 3}}}
\parskip 12pt \parindent 0pt
\hsize 16cm 
\vsize 24cm

\def\e{{\rm e}} \def\fcc{fcc} \def\hcp{hcp} \def\bcc{bcc}
\def\s{\sigma} \def\sp{\sigma^{\prime}} \def\d{{\rm d}}  
\def\br{{\bf r}} \def\rp{{\bf r}^{\prime}}
\def\SD{Spanjaard and Desjonqu\`eres}
\def\dos{g}
\def\muB{\mu\!_{_{\scriptscriptstyle B}}}
\def\eF{\varepsilon\!_F}
\def\R{{\bf R}} \def\Rp{{\bf R}^{\prime}}
\def\Rpp{{\bf R}^{\prime\prime}}
\def\L{{L}} \def\Lp{{L}^{\prime}}
\def\Lpp{{L}^{\prime\prime}}
\def\Q{Q_{\R\L}} \def\Qp{Q_{\Rp\Lp}} \def\Qpp{Q_{\Rpp\Lpp}}
\def\Qz{Q_{\R 0}} \def\Qzp{Q_{\Rp 0}} \def\Qzpp{Q_{\Rpp 0}}
\def\M{\Delta_{\ell\ell^{\prime}\ell^{\prime\prime}}}
\def\gaunt{C_{\L\Lp\Lpp}}
\def\k{{\bf k}}
\def\C{C^{n\k}_{\R\L}} \def\Cp{C^{n\k}_{\Rp\Lp}}
\def\Cb{{\bar C}^{n\k}_{\R\L}} 
\def\Cp{{\bar C}^{n\k}_{\Rp\Lp}}
\def\Sk{S^{\k}_{\R\L\Rp\Lp}}
\def\Ok{O^{\k}_{\R\L\Rp\Lp}}
\def\H{H} \def\Hin{\H_0} \def\Hp{\H^{\prime}}
\def\Etot{E_{\hbox{\tiny tot}}}
\def\Erep{E_{\hbox{\tiny pair}}} \def\Etwo{E_2^U}
\def\ro{\rho} \def\rhoin{\ro_{\hbox{\tiny in}}}
\def\rup{\ro^{+}} \def\rdn{\ro^{-}}
\def\rinup{\rhoin^{+}} \def\rindn{\rhoin^{-}}
\def\min{m_{\hbox{\tiny in}}}
\def\dro{\delta\ro} \def\droup{\dro^{+}} \def\drodn{\dro^{-}}
\def\dm{\delta m}
\def\Veff{V^\s_{\hbox{\tiny eff}}} 
\def\Vxc{V^\s_{\hbox{\tiny xc}}} 
\def\Vxcup{V^{+}_{\hbox{\tiny xc}}}
\def\Vxcdn{V^{-}_{\hbox{\tiny xc}}} 
\def\EZZ{E_{\hbox{\tiny ZZ}}}
\def\EH{E_{\hbox{\tiny H}}}  \def\EHF{E_{\hbox{\tiny HF}}}  
\def\EHin{E_{\hbox{\tiny H}}^{\hbox{\tiny in}}}
\def\Exc{E_{\hbox{\tiny xc}}}  
\def\Excin{E_{\hbox{\tiny xc}}^{\hbox{\tiny in}}}
\def\VH{V_{\hbox{\tiny H}}} \def\Vext{V_{\hbox{\tiny ext}}}
\def\EHKS{E^{\hbox{\tiny HKS}}}
\def\Eband{E_{\hbox {\tiny band}}} \def\Emag{E_{\hbox {\tiny mag}}}
\def\Veffin{V_{\hbox{\tiny eff}}^{\hbox{\tiny in}}}
\def\Vxcin{V_{\hbox{\tiny xc}}^{\hbox{\tiny in}}}
\def\Vxcs{V_{\hbox{\tiny xc}}^{\s}}
\def\VHin{V_{\hbox{\tiny H}}^{\hbox{\tiny in}}}
\def\EU{E^{\scriptstyle U}} 
\def\LDAU{LDA$+U$}
\def\Vup{V^{+}} \def\Vdn{V^{-}}

\title{Magnetic tight-binding and the iron--chromium enthalpy anomaly}

\author{A. T. Paxton}
\affiliation{Atomistic Simulation Centre, School of Mathematics and Physics,
Queen's University  Belfast, Belfast BT7 1NN, UK}  
\author{M. W. Finnis}
\affiliation{Department of Materials, Imperial College London,
Exhibition Road, London SW7 2AZ, UK} 



\begin{abstract}

We describe a self consistent magnetic tight-binding theory based
in an expansion of the Hohenberg--Kohn density functional to {\it
second order}, about a non spin polarised reference density. We
show how a {\it first order} expansion about a density having a
trial input magnetic moment leads to the Stoner--Slater rigid
band model. We employ a simple set of tight-binding parameters
that accurately describes electronic structure and energetics,
and show these to be transferable between first row transition
metals and their alloys. We make a number of calculations of the
electronic structure of dilute Cr impurities in Fe which we
compare with results using the local spin density
approximation. The rigid band model provides a powerful means for
interpreting complex magnetic configurations in alloys; using
this approach we are able to advance a simple and readily
understood explanation for the observed anomaly in the enthalpy
of mixing.

\end{abstract}

\maketitle

\section{Introduction}
\label{sec_Intro}

There is much subtlety connected with itinerant magnetism in
transition metals that one would nevertheless wish to capture in
a simple model. Recently an interatomic potential including
magnetism has been proposed\cite{Dudarev05} which will prove very
useful for molecular dynamics, but will not be able to describe
electronic structure effects such as the competition between
ferro- and antiferromagnetism, or the sudden collapse of the
moment in \hcp-Fe under pressure.\cite{Andersen77,Liu05,Drautz06}
There are very much greater difficulties attendant on interatomic
potentials employing a term in the energy which is linear in the
magnetic moment.\cite{Olsson05,Ackland06b} Almost certainly a
minimum requirement of a simple model is that it contains an
explicit account of the electron kinetic energy. This is because
inter-site magnetic interactions are carried by the hopping
matrix elements of the one-electron part of the Hamiltonian, not
by inter-site two-electron Coulomb integrals, and so Heisenberg
and Ising models are not appropriate to discuss itinerant
magnetism.\cite{Stoner38,Friedel69} The tight-binding
approximation on the other hand provides just such a
description;\cite{Harrison80,Pettifor95,Finnis03} in its most
economical form it becomes a {\it bond order potential\/}
recently described for transition metals by Drautz and
Pettifor.\cite{Drautz06} Whether a {\it magnetic} bond order
potential will appear remains to be seen; as we find below and as
pointed out in~[\onlinecite{Drautz06}] an accurate prediction of
some magnetic affects requires quite detailed structure in the
density of states near the Fermi level. Magnetic tight-binding
has been proposed many times using two slightly different self
consistent schemes. The first\cite{Roy77,Andersen77,Zhong93} is
based in a rigid band approximation first used by Andersen~{\it
et al.}\cite{Andersen77,Andersen85,Christensen88} in the context
of the local spin density approximation
(LSDA).\cite{Gunnarsson76} The non spin polarised density of
states is allowed to split rigidly as a result of on-site
exchange and correlation interactions and an energy functional
(equation~(\ref{Rigid_Functional}) below) is minimised. This
procedure may also be used in atomistic simulation if applied to
the {\it local} density of states site by site; and provides a
simple way to include effects such as magnetic pressure at
crystal defects and site dependent magnetic
moments.\cite{Yesilleten98} A second more general
approach is a self consistent scheme in which the rigid band
approximation is lifted and both the density of states and the
exchange splitting are determined self consistently.\cite{Liu05}
We are motivated to recast this procedure into our recently
proposed self consistent polarisable ion tight-binding
model,\cite{Finnis98,Fabris00,Finnis03} based on an expansion of
the Hohenberg--Kohn functional\cite{Hohenberg64} to second order
in a reference electron density. We will employ a non spin
polarised input density, which may seem surprising but is
consistent with the Stoner form of the LSDA which expands the
exchange correlation potential to linear order in the magnetic
moment.\cite{Gunnarsson76,Andersen85,Christensen88}

Having described magnetic tight-binding from the point of view of
the second order expansion, we construct very simple tight
binding models for Cr, Fe and Co which we expect to be
transferable to other transition metals and their alloys. Finally
we address an outstanding question in the thermodynamics of
Fe--Cr alloys, namely the anomalous negative enthalpy of mixing
at the Fe-rich end of the phase diagram.\cite{Mirebeau84} It is
now well known that whereas over most of the concentration range
Fe and Cr are immiscible,\cite{Hyde95} at low concentrations Cr
is soluble in Fe, with a negative enthalpy of mixing. An
explanation based on a phenomenological Ising model has been
proposed,\cite{Ackland06a} and a classical potential has been
fitted to reproduce the phase diagram.\cite{Caro05} Recent LSDA
calculations\cite{Klaver06} revealed that Cr atoms favour
clustering except at low concentrations when there is a repulsive
interaction between Cr impurities. Klaver {\it et
al.}\cite{Klaver06} pointed to this repulsive interaction in
order to explain the negative to positive upturn in the enthalpy
of mixing at concentrations in the range 8--12~atomic percent
Cr. Bandstructure arguments have been put forward based on
densities of states within the coherent potential
approximation,\cite{Olsson06} but these were rather far removed
from the actual densities, somewhat invalidating the conclusions.
We are able to advance explanations for these phenomena using
tight-binding calculations which are remarkably close to our LSDA
results and which give rise to a ready explanation easily
understood within the rigid band Stoner--Slater picture of
itinerant magnetism.

The structure of the paper is as follows. In
section~\ref{sec_SCTB} we describe how to include spin
polarisation into the self consistent polarisable ion
tight-binding model; and we decribe how the rigid band
Stoner--Slater picture may be recovered from the same framework in
section~\ref{sec_RB-Stoner}. In section~\ref{sec_model} we deduce
parameters for a simple, transferable, non orthogonal
tight-binding model for transition metals. We apply this model to
pure Fe and Cr in section~\ref{sec_FM-AFM} and to Co in
section~\ref{sec_Co}. In section~\ref{sec_Fe} we apply the model
to structural energetics of pure Fe. In section~\ref{FeCr} we
address the electronic structure of Fe--Cr alloys and in
section~\ref{sec_RB-applications} describe the use of the self
consistent rigid band model to predict the magnetic structure and
energy. We propose an explanation of the enthalpy anomaly in
section~\ref{FeCr-Enthalpy}, and conclude in
section~\ref{conclusions}. In Appendix~\ref{App_A} we show how an
equivalent form of the electron--electron interaction energy to
that derived in section~\ref{sec_SCTB} may be obtained from a
multiband Hubbard model as used in \LDAU\ theory, which exposes
the neglect of self interaction correction in LSDA and our
magnetic tight-binding while indicating how this could be put
back into a tight-binding scheme. In Appendix~\ref{App_B} we
describe non orthogonal self consistent tight-binding; in
particular we show that in this case self consistency leads to
adjustment of the {\it hopping integrals} in addition to the
on-site increments, and we illustrate the origin of additional
contributions to the interatomic force arising from bond charges.

\section{Self consistent tight-binding including magnetism}
\label{sec_SCTB}

In our self consistent polarisable ion tight-binding model we
express the electron Hamiltonian as
$$\H=\Hin+\Hp.$$ The first term is the usual non self consistent
tight-binding Hamiltonian of non interacting
electrons.\cite{Harrison80} $\Hp$ describes electron--electron
interactions and is constructed so as to represent second order
terms in the expansion of the Hohenberg--Kohn density functional
about a reference density $\rhoin$.\cite{Finnis03} We take it that
$\rhoin$ is constructed by overlapping spherical, neutral, non
spin polarised atomic charge densities. $\Hin$ is then the
Hamiltonian whose effective potential is generated by
$\rhoin$.\cite{Finnis03} We introduce a spin density
$\ro=\sum_{\s}\hbox{Tr}\hat\ro^{\s}
=\sum_{\s}\ro^{\s}=\rup+\rdn$, the electron spin taking the value
$\s=\pm 1$ in units of $\frac{1}{2}\hbar$. Minimisation of the
Hohenberg--Kohn functional leads to two Kohn--Sham
equations,\cite{Kohn65} in atomic Rydberg units,
$$\left(-\nabla^2+\Veff\right)\psi^\s=\varepsilon\psi^\s$$
in an effective potential
$$\Veff=\Vxc+\VH+\Vext$$
where $\VH$ is the Hartree potential, $\Vext$ the external
potential due to the ions and
\begin{equation}
\Vxc=\frac{\delta\Exc}{\delta\ro^\s}
\label{def_Vxc}
\end{equation}
is the exchange and correlation potential. In the absence of a
magnetic field (which we could include as a Zeeman term in
$\Vext$) this is the only term which is spin dependent. The
corresponding Hohenberg--Kohn--Sham energy functional is
(we may supress the symbol $\d\br$ under an integral sign)
\begin{align*}
\EHKS&=\sum_{{\s, n\k}\atop{{\hbox{\tiny occ.}}}}
  \left\langle\psi_{n\k}^\s\left\vert\hat T+
  \Veff\right\vert\psi_{n\k}^\s\right\rangle \\
  &-\sum_{\s}\int\ro^\s\Vxc \\ 
  &-\EH+\Exc\left[\rup,\rdn\right]
  + \EZZ \\
\end{align*}
in which $\hat T$ is the kinetic energy operator, $\EH$ is the
Hartree energy and $\EZZ$ is the ion--ion interaction.
This is expanded about the reference non spin polarised densities
$$\rinup=\rindn=\frac{1}{2}\rhoin$$
and we define
$$\dro^\s=\ro^\s-\rhoin^\s;\hskip 12pt \dro=\ro-\rhoin=\droup+\drodn.$$
The exchange and correlation energy is expanded to second order
in $\dro^\s$ to give
\begin{align*}
\Exc\left[\rup,\rdn\right]&=\Excin+\sum_{\s}\int\Vxcin\dro^\s \\
&+\frac{1}{2}\sum_{\s\sp}\int\!\!\!\int\dro^\s
     \frac{\delta^2\Exc}{\dro^\s\dro^{\sp}}\dro^{\sp}+\dots \\
\end{align*}
The Hohenberg--Kohn total energy, exact apart from the neglect of
terms higher than second order in $\Exc$ is\cite{Kohler01}
\begin{align}
E^{\scriptstyle (2)} &=
\sum_{{\s, n\k}\atop{{\hbox{\tiny occ.}}}}
\left\langle\psi_{n\k}^\s\left\vert\Hin\right\vert\psi_{n\k}^\s\right\rangle
\nonumber \\
 &-\int\rhoin\Vxcin-\EHin+\Excin+\EZZ \nonumber \\
 &+\frac{1}{2}\int\!\d\br\int\!\d\rp\left\lbrace e^2\>
   \frac{\dro(\br)\dro(\rp)}{\left\vert\br-\rp\right\vert} \right .\nonumber \\
 &+\left . \sum_{\s\sp}\dro^\s(\br)\frac{\delta^2\Exc}{\dro^\s(\br)\dro^{\sp}(\rp)}\dro^{\sp}(\rp)
               \right\rbrace. \label{E2} \\ \nonumber 
\end{align}
The first two lines amount to the
Harris--Foulkes functional.\cite{Harris85,Foulkes89,Finnis03} The
second line is represented by a pairwise repulsive energy,
$\Erep$, in the usual tight-binding models. In our self consistent
polarisable ion tight-binding model we approximate the third line
as the electrostatic interaction energy between point multipole
moments of the charge transfer. The fourth line is the extension
of the on-site electron--electron interaction Hubbard term to the
spin polarised case, and we now examine this term in more detail
using~(\ref{def_Vxc}) by writing
$$\Etwo=\frac{1}{2}\sum_{\s\sp}\int\!\!\!\int\dro^\s\>
                 \frac{\delta\Vxc}{\dro^{\sp}}\>
                 \dro^{\sp}.
$$
Here we have supressed the $\br$-dependence, firstly because all
off-diagonal Coulomb terms are relegated to the Madelung energy
(the third line in equation~(\ref{E2})) in our tight-binding
model, recognising that itinerant magnetism is a consequence of
{\it on-site} exchange and correlation;\cite{Slater36a} and
secondly because in our tight-binding model we will be using a
local orbital basis to represent the spin density. 

The quantity
\begin{equation}
\frac{\delta\Vxcup}{\drodn}=\frac{\delta\Vxcdn}{\droup}\equiv U
\label{def_Vupdn}
\end{equation}
is the direct Coulomb, correlation only, interaction strength between
unlike spins described by the Hubbard $U$ parameter. On the other
hand the quantity
\begin{equation}
\frac{\delta\Vxcup}{\droup}=\frac{\delta\Vxcdn}{\drodn}\equiv
U-I
\label{def_Vupup}
\end{equation}
reflects the lowering of the electron--electron interaction
through exchange by an amount $I$, here called the Stoner
parameter. Because of the Pauli principle electrons with like
spins are kept further apart and so their electrostatic Coulomb
repulsion is, on average, weaker than for unlike spin
electrons. This is the origin of Hund's rule as well as spin
polarisation of itinerant electrons. Using these definitions of
$U$ and $I$ we can write down $\Etwo$ in terms of the total
density and the magnetic moment (equation~(\ref{Eq_E2})
below). First, we note that the magnetic moment $m$ is
$$
m=\rup-\rdn=\droup-\drodn=\dm
$$
since the input density is non spin polarised. We then find, 
using~(\ref{def_Vxc}),~(\ref{def_Vupdn}) and~(\ref{def_Vupup})
$$\frac{\delta^2\Exc}{\delta\ro^2}=U-\frac{1}{2}I,$$
whereas\cite{Shimizu64,Footnote4}\nocite{Janak77}
$$I=-2\>\frac{\delta^2\Exc}{\delta m^2},$$
where the second derivatives are to be evaluated at the input
density, {\it i.e.,} $m=0$.

We also have,
$$\droup=\frac{1}{2}\left(\dro+\dm\right),\hskip 12pt
  \drodn=\frac{1}{2}\left(\dro-\dm\right)$$
from which we readily obtain the central result of this section,
\begin{equation}
\Etwo=\frac{1}{2}U\dro^2-\frac{1}{4}I\dro^2-\frac{1}{4}Im^2.
\label{Eq_E2}
\end{equation}
Only the first two terms survive in the non spin polarised model
described previously.\cite{Finnis98,Fabris00,Finnis03} An
associated expression may be obtained from the \LDAU\
formalism as demonstrated in Appendix~\ref{App_A}.
Finally, we give the expression for the tight-binding total
energy including the magnetic terms,
\begin{equation}
\Etot= E_1 + E_2
\label{etot}
\end{equation}
with
$$E_1=\sum_{\s}\hbox{Tr}\left[\hat\rho^{\s}\Hin\right]+\Erep$$
and
\begin{align*}
E_2 &= \frac{1}{2}\sum_{\R}\left\lbrace\sum_{\L}\Q V^M_{\R\L}\right. \\
    &+ \left.\left(U_{\R}-\frac{1}{2}I_{\R}\right)\delta q_{\R}^2
        -\frac{1}{4}I_{\R}m_{\R}^2\right\rbrace \\
\end{align*}
in which $\hat\rho^{\s}$ is the spin density matrix,
$\R$ labels atomic sites and $\delta q_{\R}$ and
$V^M_{\R\L}$ are as defined in equations~(\ref{def_q})
and~(\ref{def_V}) in Appendix~\ref{App_B}. 
There are no additional contributions to the interatomic force
due to spin polarisation.\cite{Liu05} 

\section{Rigid band Stoner--Slater model}
\label{sec_RB-Stoner}

\begin{figure}
\begin{center}
\includegraphics[scale=0.4,angle=0, trim=0 0 0 0]{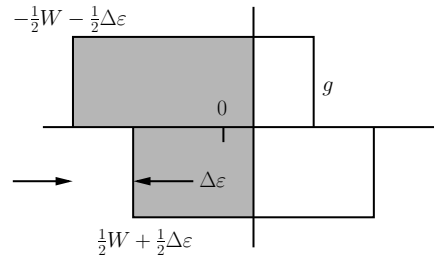}
\end{center}
\caption{To illustrate the simple rectangular density of states
model of ferromagnetism. A density of states which is constant
and equal to $g$ between band edges $\pm\frac{1}{2} W$ is split
by exchange into majority and minority spin densities by an
amount proportional to the moment, $m$. (This can be achieved by
flipping the spins of $\frac{1}{2} m$ electrons and realigning
the Fermi levels to a common value.) By construction, we have
$\Delta\varepsilon=m/g=Im$.  (After Pettifor, \cite{Pettifor95}
figure~8.12c)}
\label{Stoner}
\end{figure}

In the previous section we expanded the Hohenberg--Kohn total
energy to second order around a non spin polarised reference
density. Alternatively one may expand about a spin polarised
density having a non zero trial magnetic moment.\cite{Pickett96}
We now show that in this case an expansion to {\it first order}
is appropriate and that the resulting Harris--Foulkes functional
may lead to the well known rigid band Stoner--Slater
model,\cite{Stoner33,Stoner36,Slater36a,Slater36b} (usually
referred to as just the ``Stoner model''). We recall first that
this is most readily
illustrated\cite{Friedel64,Pettifor80,Pettifor95} using the
rectangular density of states, representing the $d$-band in a
transition metal shown in figure~\ref{Stoner}. We imagine that
majority spin electrons see an exchange and correlation potential
lower than that seen by minority electrons by an amount
proportional to the magnetic moment, $m$; the proportionality
constant, $I$, being the ``Stoner parameter.'' (Stoner uses the
symbol $\alpha$ for this, $I$ is Slater's usage.\cite{Slater36a})
Then the rectangular bands are split by
$\pm\frac{1}{2}\Delta\varepsilon=\pm\frac{1}{2}Im$ and the change
in band (kinetic) energy due to magnetisation is
\begin{align*}
\Delta\Eband = {\hskip 36pt} {} & \\
\left(
\int_{-\frac{1}{2} W-\frac{1}{2}\Delta\varepsilon}^{\eF} 
  \hskip -12pt g\varepsilon\d\varepsilon
\right.
 &+\left.\int_{-\frac{1}{2} W+\frac{1}{2}\Delta\varepsilon}^{\eF}
  \hskip -12pt   g\varepsilon\d\varepsilon
 -2\int_{-\frac{1}{2} W}^{\eF}  \hskip -12pt g\varepsilon\d\varepsilon
\right)
\\
 {\hskip -36pt}  &= -\frac{1}{4}gI^2m^2,
\end{align*}
using $\Delta\varepsilon=Im$. In this estimate of the magnetic
energy the electron--electron interaction energy,
$-\frac{1}{4}Im^2$ has been double counted, so it is subtracted
to give
$$\Delta \Emag=\frac{1}{4}Im^2\left(1-Ig\right)$$ which is
negative as long as $Ig>1$, which is the simplest statement of
the Stoner criterion.\cite{Stoner33} This particular model is
pathological because $\Delta\Emag$ has no minimum as a function
of $m$. This is a symptom of using a constant density of states,
so that the kinetic energy is quadratic in $m$; that is, the
fourth order term which is responsible for stabilising the
ferromagnetic state is missing in the absence of structure in the
density of states.

According to Slater,\cite{Slater36a} ferromagnetism
arises from a competition between kinetic energy and {\it
on-site} Coulomb electron--electron interactions. For an
arbitrarily shaped density of states the kinetic energy increases
compared to the spin-paired state when down-spin electrons are
spin-flipped, since they must then be promoted into unoccupied
states above the Fermi level. To develop a magnetic moment, $m$,
charge is transferred across the Fermi surface in small
increments $\d m$, each increment costing more energy than the
last as the down-spin states are depleted below the Fermi level
and need to be taken from lower energy states and placed as
up-spin electrons in higher energy states as these become
successively occupied above the Fermi level. Generally speaking
the larger the density of states near the Fermi level the smaller
is the energy penalty involved. To counter this increase in
kinetic energy there will be a decrease in energy due to a Hund's
rule like exchange interaction and Slater\cite{Slater36a} argues
that this takes the form $-\frac{1}{4}Im^2$. Hence the total
change in energy upon forming a magnetic moment $m$
is\cite{Friedel64,Shimizu64,Roy77,Andersen77,Christensen88}
\begin{equation}
\Delta \Emag(m)=\frac{1}{2}\int_0^m\frac{m'\d m'}{\bar g(m')}-\frac{1}{4}Im^2,
\label{Rigid_Functional}
\end{equation}
which is clearly stationary at a generalised Stoner condition,
namely $I\bar g(m)=1$,
where $\bar g(m)$ is the density of states averaged over the
energy range spanned by flipping the $\frac{1}{2}m$ spins; see
figure~30 in the Varenna notes.\cite{Andersen85}

This is a rigid band model, requiring us to know only the {\it
non magnetic} density of states. We can obtain an analogous
expression for $\Delta\Emag$ from a Harris--Foulkes functional,
namely the first two lines of equation~(\ref{E2}). In contrast to
the second order theory in which the input density is non spin
polarised, let us consider a trial density which can be varied by
changing its magnetic moment while not affecting the total charge
density.\cite{Pickett96} We now have $\min=\rinup-\rindn$ and the
trial Hamiltonian is 
$$H^{\s}=\Hin+\Vxcs[\rhoin]$$
where
$$\Vxcs[\rhoin]=-\frac{1}{2}\s I\min$$
so that $\s=+1$ are the majority spins (\ie, see a lower exchange
and correlation potential). We now evaluate the first order total
energy, 
\begin{equation*}
\begin{split}
E^{\scriptstyle (1)}(\min) &=
\sum_{{\s, n\k}\atop{{\hbox{\tiny occ.}}}}
\left\langle\psi_{n\k}^\s\left\vert\H^{\s}\right\vert\psi_{n\k}^\s\right\rangle\\
 &-\sum_{\s}\int\rhoin^{\s}\Vxcs[\rhoin]-\EHin+\Excin+\EZZ. \\
\end{split}
\end{equation*}
When we compare this to its value when $\min=0$ we
obtain
\begin{equation}
\Delta \Emag(\min)=\Delta\Eband(\min) + \frac{1}{4}I\min^2
\label{emag}
\end{equation}
after evaluating the double counting in view of the fact that
only the moment and not the density differ in the two cases, and
using $\Delta\Excin=-\frac{1}{4}I\min^2$.
\nocite{Footnote6,VonBarth72,Moruzzi78} 

\begin{figure}
\begin{center}
\includegraphics[scale=0.4,angle=0, trim=0 0 0 0]{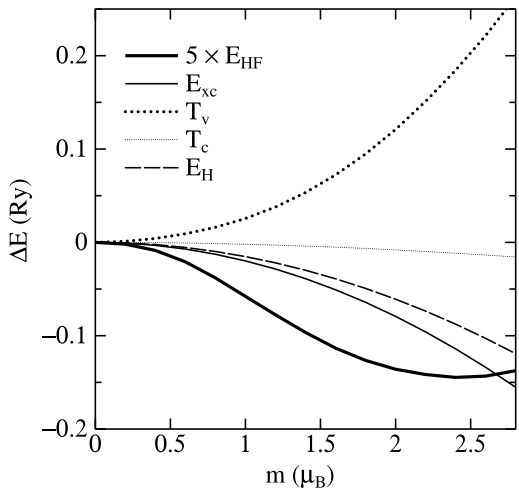}
\end{center}
\caption{Contributions to the total energy in an
LMTO\protect\cite{Footnote6} calculation for
pure \bcc-Fe relative to their values at $m=0$. This is the
Harris--Foulkes energy,\protect\cite{Harris85,Foulkes89} $\EHF$, as a
function of the fixed magnetic moment of the input density,
$\rhoin$. $T_c$ and $T_v$ are core and valence kinetic energies
and $\EH$ is the Hartree energy. We find that $\Exc$ is almost
exactly quadratic and hence its curvature is independent of
$m$. Its curvature here is $-0.04$, giving $I=80$~mRy compared to
the value 65~mRy\protect\cite{Poulsen76,Mackintosh80,Andersen85} using
both Janak's method\protect\cite{Janak77} and that of Poulsen {\it et
al.}\protect\cite{Poulsen76} and 68~mRy as calculated by
Gunnarsson.\protect\cite{Gunnarsson76} The fourth order term in $\EHF$
which leads to a minimum at the observed moment comes from the
kinetic energy.}
\label{Stoner-LSDA}
\end{figure}
As an illustration, we show in Figure~\ref{Stoner-LSDA} how a
Harris--Foulkes energy varies with moment in pure \bcc-Fe. Here,
we have constructed an input density by superimposing free
atoms\cite{Polatoglou90,Paxton90} having a given magnetic moment
so that the moment of the input density is a trial $\min$. We
then evaluate the Harris--Foulkes total energy functional and
plot it against $\min$. This is not exactly a rigid band
calculation, but it serves to illustrate how the individual
contributions to the energy vary with $\min$. In particular note
that the kinetic energy increases, having both second and fourth
order terms in $\min$, while the exchange and correlation energy
is found to be strictly quadratic. This is consistent with the
Stoner picture and serves to show that the Stoner parameter $I$
is independent of the moment and so may be taken as the same
quantity in both equations~(\ref{etot}) and~(\ref{emag}). Our
estimate of $I$ is of course not as good as a fully self
consistent calculation as we indicate in the caption to
figure~\ref{Stoner-LSDA}.

We will use equation~(\ref{etot}) to calculate density of states
and total energy in sections~\ref{sec_FM-AFM} to~\ref{FeCr}. The
rigid band picture is particularly useful in interpreting complex
magnetic structures and arriving at an explanation of the
enthalpy anomaly. Therefore in sections~\ref{sec_RB-applications}
and~\ref{FeCr-Enthalpy} we employ equation~(\ref{emag}) to find
the total energy.

\section{Tight-binding model}
\label{sec_model}

Our tight-binding model is specified by distance dependent matrix
elements of the Hamiltonian and overlap, by Hubbard~$U$ and
Stoner~$I$ parameters, and by a repulsive pair potential. We are
motivated to employ the simplest possible scheme so as to
maximise its predictive power relative to its
complexity.\cite{Finnis03} Our starting point is the
tight-binding theory of transition metals of \SD\cite{SD} who
propose a universal, orthogonal scheme in which Hamiltonian
matrix elements have the form $f_0\e^{-qd}$ and the pair
potential takes the form $B\e^{-pd}$, where $d$ is the bond
length. These are intended to extend to nearest neighbours only
in \fcc\ and \hcp\ metals and to second neighbours in the \bcc\
structure. \SD\ find a universal ratio $p/q=2.95$ that fits well
to the binding energy curve of Rose~\ea\cite{Rose84} We have
found this to be an excellent model for transition metals using
an orthogonal basis of $d$-electrons\cite{Paxton96} and adopting
the canonical ratio for the three quantities $f_0$, namely
$$dd\s:dd\pi:dd\delta = -6f_0:4f_0:-1f_0.$$
\SD\ provide values of the product $qd_0$, where $d_0$ is the
equilibrium bond length, for most transition metals. Therefore the
only adjustable parameters are $f_0$ which we adjust to
the bandwidth calculated in the LDA, and the parameter $B$
which we adjust to obtain the correct atomic volume (or lattice
constant). This simple model having two adjustable parameters
then gives a good account of structural stability and elastic
constants.\cite{Paxton96}

For a number of reasons, we wish to go beyond this very simple
scheme in three respects. ({\it i\/}) We will extend the range of
the exponentially decaying interactions; specifically we
encompass 58 neighbours in the \bcc\ lattice. This has the
attraction of employing an energy surface without discontinuities
in a molecular dynamics simulation. Furthermore we have found
this necessary to obtain a faithful reproduction of the LDA
density of states. ({\it ii\/}) For this latter reason we also
prefer to include $s$ and $p$ electrons in the basis, and ({\it
iii\/}) to adopt a non orthogonal basis. We see a number of
attractions from the inclusion of overlap which we discuss in
Appendix~\ref{App_B} (see also the caption to
figure~\ref{FM-Fe-dos}, below). It is furthermore known that the
neglect of $sd$-hybridisation leads to an overestimation of the
magnetic moment of Fe.\cite{Poulsen76,Pettifor83,Yesilleten98}
Our procedure for obtaining the additional parameters is again
motivated by simplicity and we adjusted the additional matrix
elements to obtain a close comparison between the LDA and
tight-binding density of states in \bcc\ Fe. Thereafter we merely
adjusted $f_0$ to allow for the differences in $d$-bandwidth
across the transition series. We use the same exponent in the
overlap matrix elements as in the Hamiltonian, but with a
different prefactor, they thereby take the form $s_0\e^{-qd}$. We
use $qd_0^{\bcc}=3$ for all $dd$ interactions otherwise we set
$q=0.5$~bohr$^{-1}$. We deviated from the canonical ratios in the
non orthogonal case:
$$dd\s:dd\pi:dd\delta = -6f_0:5f_0:-2.2f_0,$$
and furthermore used the ratio
$$pp\s:pp\pi=2:-1.$$ We fix the on-site energy levels of the $s$
and $p$ atomic levels at 0.2~Ry and 0.45~Ry respectively,
relative to the $d$-level.  The remaining parameters are shown in
table~\ref{tbl_parms}.

\begin{table*}
\caption{Parameters of our tight-binding model. Atomic Rydberg
units are used throughout.}
\begin{tabular}{|l @{\hskip 4pt}| c @{\hskip 4pt} c @{\hskip 4pt}| c @{\hskip 4pt} c @{\hskip 4pt}| c @{\hskip 4pt} c @{\hskip 4pt}| c @{\hskip 4pt} c @{\hskip 4pt}| c @{\hskip 4pt} c @{\hskip 4pt}| c @{\hskip 4pt} c @{\hskip 4pt}| c @{\hskip 4pt}| c|}\hline
 & \multicolumn{2}{c|}{$ss\s$}
 & \multicolumn{2}{c|}{$sp\s$}
 & \multicolumn{2}{c|}{$pp\s$}
 & \multicolumn{2}{c|}{$sd\s$}
 & \multicolumn{2}{c|}{$pd\s$}
 & \multicolumn{2}{c|}{$dd\s$}
 & $B$ & $I$\\
 & $f_0$ & $s_0$  & $f_0$ & $s_0$  & $f_0$ & $s_0$  & $f_0$ & $s_0$  &
 $f_0$ & $s_0$  & $f_0$ & $s_0$ &  &\\
\hline\hline
Cr & --0.75 & 0.5 & 0.5 & --1.0 & 1.0 & --0.1 & --0.12 & 0.8 &
 --0.5 & 0 & 0.18 & 0 & --- & 0.050\\
Fe & --0.75 & 0.5 & 0.5 & --1.0 & 1.0 & --0.1 & --0.12 & 0.8 &
 --0.5 & 0 & 0.12 & 0 & 340 & 0.055\\
Co & --0.75 & 0.5 & 0.5 & --1.0 & 1.0 & --0.1 & --0.12 & 0.8 &
 --0.5 & 0 & 0.10 & 0 & 250 & 0.080\\
\hline
\end{tabular}
\label{tbl_parms}
\end{table*}
Our values of the Stoner $I$ are essentially those calculated by
Gunnarsson and
others.\cite{Gunnarsson76,Poulsen76,Janak77,Mackintosh80,Andersen85}
However we adjust these to obtain magnetic moments in agreement
with the LSDA.

\begin{figure}
\begin{center}
\includegraphics[scale=0.4,angle=0, trim=0 0 0 0]{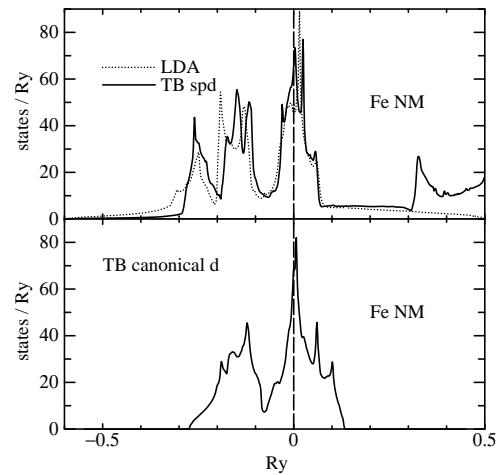}
\end{center}
\caption{Density of states of non magnetic Fe using two
tight-binding models. The upper panel shows the non orthogonal model
having the parameters shown in table~\ref{tbl_parms}; the dotted
line shows the LDA density of states. The lower panel shows the
density of states in the canonical $d$-band tight-binding model.}
\label{NM-Fe-dos}
\end{figure}

Figure~\ref{NM-Fe-dos} illustrates the match between LDA and
tight-binding densities of states in the non orthogonal and
orthogonal $d$-only tight-binding models. Note that the canonical
model is quite adequate in describing the essential features,
namely the $t_{2g}$ ($xy$, $yz$, $zx$) bonding and $e_g$
($x^2-y^2$, $z^2-r^2$) antibonding manifolds which stabilise the
\bcc\ structure at half band filling and the large density of
states at the Fermi level, $\dos(\eF)$, which is responsible for
the ferromagnetic instability. To place the Fermi level exactly
at the peak, it is necessary to choose the number of
$d$-electrons, $N_d$, as an additional parameter in the $d$-only
tight-binding model; we set $N_d=6$. However the three peak
structure typical of \bcc\ transition metals and the smooth
``U''-shaped pseudogap are less faithfully reproduced in the
canonical model.

\begin{figure}
\begin{center}
\includegraphics[scale=0.4,angle=0, trim=0 0 0 0]{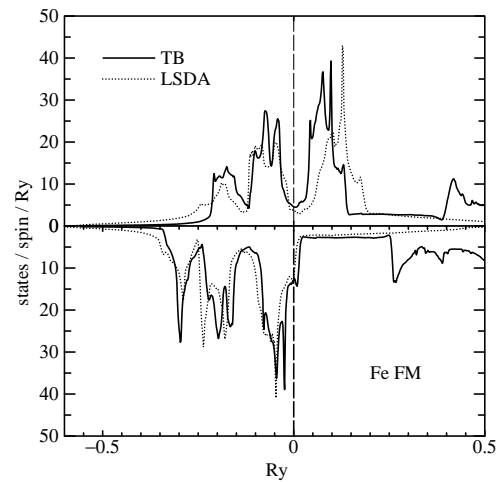}
\end{center}
\caption{Density of states in ferromagnetic Fe using the $spd$
tight-binding model and compared to an LSDA calculation. The
upper panel shows the minority and the lower panel the majority
spins. Note that in an {\it orthogonal} tight-binding model, even
using the fully self consistent scheme of section~\ref{sec_SCTB},
the two densities of states would be identical, only rigidly
shifted. The inclusion of an overlap breaks this symmetry and it
is seen here that this additional freedom acts significantly to
improve the comparision with the LSDA.}
\label{FM-Fe-dos}
\end{figure}

\begin{figure}
\begin{center}
\includegraphics[scale=0.4,angle=0, trim=0 0 0 0]{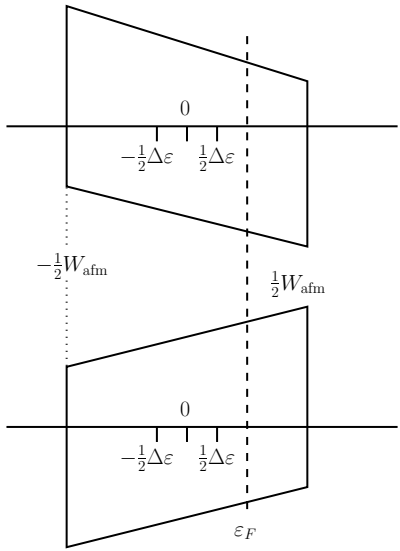}
\end{center}
\caption{Illustration of Pettifor's skewed rectangular band model
of antiferromagnetism. The upper and lower figures show the
densities of states on the two sublattices. A single one of
these describes the situation in an AB alloy in which the
electrons see a lower potential, say, at the A-atom whose density
of states is accordingly skewed toward lower eigenvalues as in
the upper density of states of the top diagram. In the
antiferromagnetic analogy, on each sub lattice the majority spin
electrons see a lower potential due to exchange interactions,
they spend more time at that site and the density of states is
accordingly skewed. Spin up are the majority electrons at one
sublattce, spin down at the other; hence the two diagrams, one
for each sublattice. (After Pettifor,\cite{Pettifor95}
figure~8.12b)}
\label{Skew}
\end{figure}

\section{Ferro- and antiferromagnetism in pure Iron and Chromium}
\label{sec_FM-AFM}

\begin{figure}
\begin{center}
\includegraphics[scale=0.4,angle=0, trim=0 0 0 0]{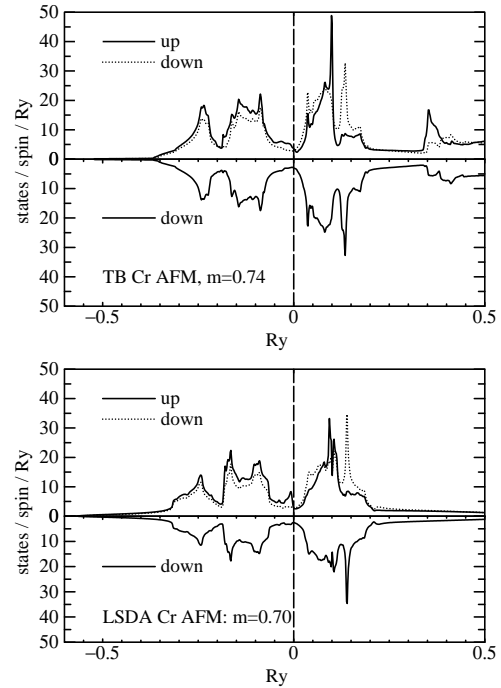}
\end{center}
\caption{Density of states in antiferromagnetic Cr showing both
the $spd$ tight-binding model and an LSDA calculation. In the
upper panels are shown both the up spins and the down spins from
the lower panel reflected in the $x$-axis to reveal the lower
density of states compared to the up spin. The reverse situation
pertains on the other sublattice. The lower density of states
over the occupied density of states is the generalisation of the
skewed rectangular band picture. }
\label{AFM-Cr-dos}
\end{figure}

It is quite clear that both canonical and $spd$ tight-binding
models predict ferromagnetism in Fe based in the Stoner
criterion, $I\dos(\eF)>1$, which in the simplest rectangular band
models of Friedel\cite{Friedel64} and Pettifor\cite{Pettifor80}
is $I/W>1/5$, where $W$ is the width of the
$d$-band.\cite{Pettifor95} In figure~\ref{FM-Fe-dos} we show the
self consistent tight-binding density of states compared to the
LSDA. We find a self consistent magnetic moment of
$2.18\muB$. The density of states of Cr is of course of
practically the same shape as that of Fe but the Fermi level
falls inside the pseudogap. In Pettifor's {\it skewed}
rectangular $d$-band theory,\cite{Pettifor80,Pettifor95}
antiferromagnetism is predicted if
$$\frac{I}{W}>\left[\frac{3}{10}N_d\left(10-N_d\right)\right]^{-1}.$$
In this theory, the analogy is made between an AB binary alloy
and an antiferromagnetic crystal having two sublattices, as does
the \bcc\ structure. In the alloy electrons will see a lower
potential, say, at the A-site where the on-site energy level is
lower than at the B-site by an amount $\Delta\varepsilon$. In the
common band model this leads to a skewing of the simple
rectangular density of states, so that lower energy eigenvalues
are generally associated with the A-site and {\it vice versa.} In
this picture electrons in the lower energy single particle states
spend more time at the A-site while overall charge neutrality is
maintained.\cite{Pettifor87} In the antiferromagnetic case
(figure~\ref{Skew}) one says that up-spin electrons see a lower
exchange potential at one sublattice and the down spin at the
other. Each of their on-site energies are lowered through the
exchange interaction (Hund's rule) by an amount
$\Delta\varepsilon=Im$, if $I$ is sufficiently large, which
favors aligned spins. Figure~\ref{AFM-Cr-dos} shows that this
effect is predicted in the self consistent tight-binding model
and compares the resulting density of states with the LSDA. The
local antiferromagnetic moment $m$ in the tight-binding model is
predicted to be $0.74\muB$ in close agreement with the $0.70\muB$
estimated from the LSDA spin density.

\begin{figure}
\begin{center}
\includegraphics[scale=0.4,angle=0, trim=0 0 0 0]{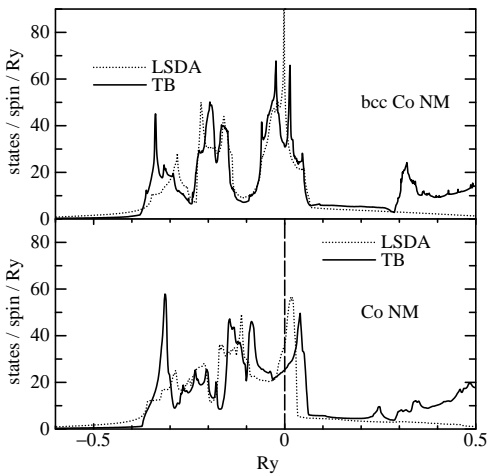}
\end{center}
\caption{Density of states in non magnetic Co: upper figure \bcc,
lower figure \hcp. Dotted lines show the LDA densities of
states. Note that the parameters generated for the \bcc\
structure transfer well to the observed \hcp\ structure.}
\label{NM-Co-dos}
\end{figure}

\section{Transferability to Cobalt}
\label{sec_Co}

We begin discussion of energetics with the application of the
\SD\ model to Co. The approach we have taken is to adjust the
parameter $f_0$ only to match the $d$-bandwidth of non magnetic
\bcc-Co calculated in the LDA. The resulting density of states is
shown in figure~\ref{NM-Co-dos} which also shows the density of
states in \hcp-Co to demonstrate the transferability of the band
parameters to the observed structure of Co.

The remaining parameter, $B$, that enters the pair potential was
fitted to the calculated lattice constant of {\it non magnetic}
\bcc-Co. Table~\ref{Co_data} shows the results of calculations of
both \bcc\ and \hcp\ Co. The model is clearly remarkably
predictive and argues strongly for the essential correctness of
the \SD\ approach coupled to the second order Stoner
theory. Particularly, note that the tight-binding correctly
predicts the stability of the \hcp\ over the \bcc\ structure and
also renders rather well the bulk moduli, both in magnetic and
non magnetic forms. In connection with the Stoner~$I$ parameter,
we note firstly that the value, 68~mRy, quoted for the LSDA is
not, of course, an input into the calculation but this is the
number calculated by other authors using the LSDA
approach.\cite{Andersen85} Secondly, we note that we tried two values
in the tight-binding model: $I=80$~mRy gives a better value of
the magnetic moment in \bcc-Co, whereas this value gives a
negative magnetic energy for \hcp-Co thus predicting this phase
to be non magnetic. Increasing $I$~to 85~mRy corrects this but
overstates the moment in \bcc-Co.

\begin{figure}
\begin{center}
\includegraphics[scale=0.4,angle=0, trim=0 0 0 0]{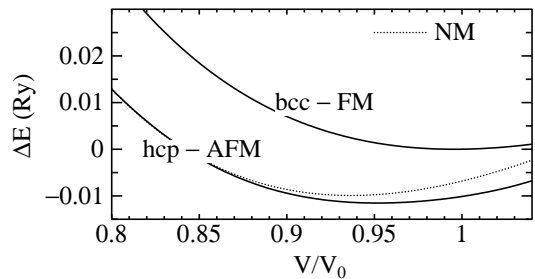}
\end{center}
\caption{Energy volume curves in Fe using the $spd$ tight-binding
model of table~\ref{tbl_parms}. We used \hcp\ having ideal axial
ratio. The dotted line shows non magnetic \hcp; the two curves
merge as the antiferromagnetic moment vanishes with reducing
atomic volume.}
\label{Fe_E-V}
\end{figure}

\begin{table*}
\caption{Energetic data for Co, comparing tight-binding and
LSDA calculations. Note that the only fitted values are the
atomic volume of \bcc-Co, although the Stoner~$I$ has also been
adjusted to agree with the LSDA moments. $V/V_0$ is the atomic
volume compared to experiment; $\Delta E_{\rm h-b}$ is the energy
of the \hcp\ relative to the \bcc\ phase; $m$ is the magnetic
moment; $\Delta\Emag$, the ``magnetic energy'' is the
calculated energy difference between magnetic and non magnetic
phases; $K$ is the bulk modulus.}

\begin{tabular}{|@{\hskip 1pt}c@{\hskip 1pt}|@{\hskip 1pt}c@{\hskip 1pt}| @{\hskip 1pt}c@{\hskip 1pt} | @{\hskip 4pt}c @{\hskip 4pt}c |c @{\hskip 4pt}c @{\hskip 4pt}|c c| c @{\hskip 4pt}c@{\hskip 4pt}c|}\hline
  & $I$ & & \multicolumn{2}{c|}{$\Delta E_{\rm h-b}$}
  &     \multicolumn{2}{c|}{$m$ ($\muB$)}
  &     \multicolumn{2}{c|}{$\Delta\Emag$}
  &     \multicolumn{3}{c|}{$K$ (Mbar)} \\
  & (mRy) & & \multicolumn{2}{c|}{(mRy)}
  & & &    \multicolumn{2}{c|}{(mRy)}
  & & &\\
 & & $V/V_0$ & TB & LSDA & TB & LSDA & TB & LSDA & TB & LSDA & expt. \\
 \hline\hline
 {bcc} & 0  & {0.896} & & & & {0 NM} & & & & {2.94} & \\
 bcc & 0 & {0.896} & & & 0 NM & & & &  {3.04} &
  & \\
 \hline
 {bcc} & {68} & {0.935} & & & & {1.67 FM} & & {18} & & {2.52} & \\
 bcc & 80 & 0.933 & & & 2.08 FM & & 18 & & 2.81 & & \\
 {bcc} & {85} & {0.935} & & & {2.16 FM} & & {22} & & {2.82} & & \\
 \hline
 {hcp} & 0 & {0.878} & & {--18}
  & & {0 NM} & & & {3.11} & &\\
 hcp & 0 & 0.875 & --30 & & 0 NM & & & & & 3.42 & \\
 \hline
 {hcp} & {68} & {0.916} & & {--12} & & {1.55 FM} & & 11 & & {2.71} & 1.91 \\
 hcp & 80 & 0.921 & --12 & & 1.91 FM & & --0.1 & & 2.90 & & \\
 {hcp} & {85} & {0.924} & {--11} & & {1.99 FM} & & {4} & & {2.92} & & \\
 \hline
\end{tabular}
\label{Co_data}
\end{table*}

\section{Phase stability in Iron}
\label{sec_Fe}

We continue to look at the energetics by examining how the simple
\SD\ model describes the stability of the close packed structures
in Fe. This has been addressed in detail recently,\cite{Liu05} so
for brevity we discuss only the \bcc\ and \hcp\ structures at two
atomic volumes, $V/V_0=1$ and 0.88 where $V_0=11.82$\r{A}$^{3}$
is the experimental atomic volume of \bcc-Fe and the transition
to \hcp-Fe is observed\cite{Clendenen64} to occur at about
$V/V_0=0.88$.  Table~\ref{Fe_data} shows that the predictions are
less accurate than in the case of Co. We recall that very careful
studies of the energetics in the LSDA have been made by Bagno
\ea\cite{Bagno89} and by Stixrude \ea\cite{Stixrude94} The
conclusions are that at $V/V_0=1$, the most stable phase is
ferromagnetic \bcc-Fe, but that the energy volume curve for
antiferromagnetic \hcp-Fe intersects that for \bcc-Fe and has a
minimum at a lower energy at $V/V_0\approx 0.88$. Hence the
global prediction of the LSDA is that \hcp\ is the stable phase
having a higher than ambient density. It is well known that this
anomaly is removed by use of the so called generalised gradient
approximation (GGA), although Bagno \ea\ point out that this is
probably merely a coincidence arising from the GGA favouring of
both larger atomic volumes and larger magnetic moments as a
general rule. As can be seen in figure~\ref{Fe_E-V}, our
tight-binding model rather closely follows the LSDA, but fails to
reproduce the stability of \bcc-Fe even at the ambient atomic
volume. Table~\ref{Fe_data} shows also the predicted magnetic
moments and bulk modulus. Note that we have used the ideal axial
$c/a$ ratio for \hcp\ at $V/V_0=1$, but its measured value at
$V/V_0=0.88$.

\begin{table*}
\caption{Energetics of Fe in the \bcc\ and \hcp\ crystal
structures. Note that the tight-binding model incorrectly
predicts that \hcp\ is stable at $V/V_0=1$, but correctly
reproduces the LSDA result that \hcp\ is stable at
$V/V_0=0.88$. The magnetic energies show that \hcp\ is only very
weakly antiferromagnetic, especially at high pressure and this
result is correctly reproduced by the tight-binding model.}
\begin{tabular}{|@{\hskip 1pt}c@{\hskip 1pt}| @{\hskip 1pt}c@{\hskip 1pt} | @{\hskip 1pt}c@{\hskip 1pt} | @{\hskip 4pt}c @{\hskip 4pt}c |c @{\hskip 4pt}c @{\hskip 4pt}c @{\hskip 4pt}|c c| c @{\hskip 4pt}c|}\hline
  & & & \multicolumn{2}{c|}{$\Delta E_{\rm h-b}$}
  &     \multicolumn{3}{c|}{$M$ ($\muB$)}
  &     \multicolumn{2}{c|}{$\Delta\Emag$}
  &     \multicolumn{2}{c|}{$K$ (Mbar)} \\
  & & & \multicolumn{2}{c|}{(mRy)}
  & & & &    \multicolumn{2}{c|}{(mRy)}
  & &\\
 & $V/V_0$ & $c/a$ & TB & LSDA & TB & LSDA & expt. & TB & LSDA &
 TB & expt. \\
\hline\hline
bcc & 1 & -- & 0 & 0 & 2.18 FM & 2.08 FM & 2.21 FM & 17 & 30 & 2.24 & 1.68 \\
hcp & 1 & 1.63 & --10 & +6 & 1.8 AFM & 1.57 AFM & & 3.2 & 1.1 & & \\
hcp & 0.88 & 1.58 & --7 & --15 & 0.9 AFM & 0.04 AFM & & $\sim 0$ & $\sim 0$ &&\\
\hline
\end{tabular}
\label{Fe_data}
\end{table*}

Maybe it is not surprising that this very simple tight-binding
model fails to describe the energetics of Fe. This is a very
subtle problem even for the LSDA. The solution within
tight-binding is rather simple however as has been demonstrated
recently, and requires the use of a more complicated pair
potential.\cite{Liu05} This is consistent with the observations
of Bagno \ea\cite{Bagno89} concerning the role of the GGA, and
need not concern us further here, since in what follows we will
discuss electronic structure and leave aside the question of
structural energetics.

\section{Electronic structure in the Iron--Chromium alloy system}
\label{FeCr}

\subsection{FeCr in the B2 crystal structure}

For the remainder of the paper we discuss the electronic
structure of Fe--Cr alloys. It is very simple to construct a
model for interactions between Fe and Cr by taking the geometric
mean of the $d$--$d$ hopping integrals and by moving the on-site
$d$-orbital energies up and down by 0.1~Ry. Thereby one would
expect a small charge transfer from Cr to Fe, since the latter is
more electronegative. To control this charge transfer we apply a
Hubbard~$U$ of 1~Ry. Our model deviates in this way slightly from
the usual {\it ansatz} of local charge
neutrality.\cite{Pettifor87}

\begin{figure}
\begin{center}
\includegraphics[scale=0.4,angle=0, trim=0 0 0 0]{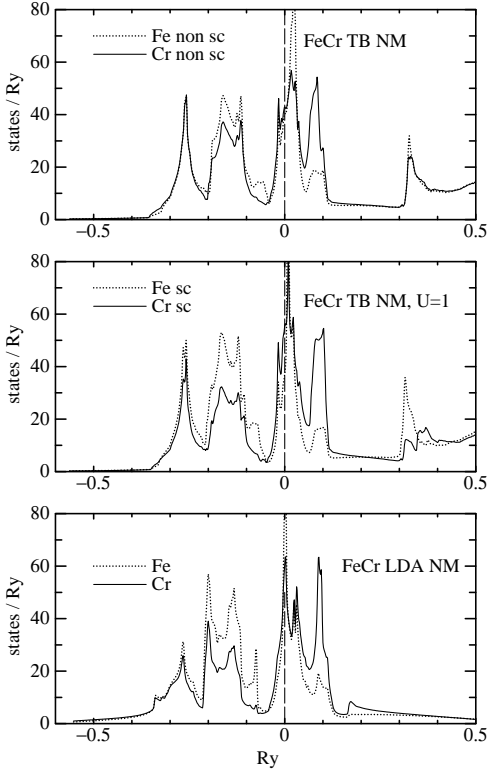}
\end{center}
\caption{Atom projected densities of states in non magnetic
FeCr. The top panel shows a non self consistent tight-binding
calculation in which the charge transfer is seen to be somewhat
smaller than that found from the self consistent tight-binding
calculation in the centre panel. This self consistent result
agrees better with the LDA in the lower panel.} 
\label{NM-FeCr-dos}
\end{figure}

The B2 alloy FeCr has a positive heat of formation and hence does
not exist.\cite{Singh94,Klaver06} Nonetheless it presents an
interesting case in which to discuss the competition between
ferro- and antiferromagnetism. One might expect this ordered alloy
to be antiferromagnetic since the Cr sublattice could prefer to
align antiferromagnetically with the neighbouring Fe atoms. But
the non magnetic density of states clearly shows a large density
of states at the Fermi level and one expects the Stoner criterion
to apply and lead to ferromagnetism. However, it turns out in the
tight-binding model that both ferro- and antiferromagnetic
solutions can be found depending on the value of the Hubbard~$U$;
but in the physically correct limit of large~$U$ the alloy is
ferromagnetic in agreement with the LSDA. To begin with,
figure~\ref{NM-FeCr-dos} shows Fe and Cr atom projected densities
of states in non magnetic FeCr. We observe that the small amount
of charge transfer permitted by the self consistent tight-binding
leads to a closer agreement with the LDA than the non self
consistent tight-binding density of states.

\begin{figure}
\begin{center}
\includegraphics[scale=0.4,angle=0, trim=0 0 0 0]{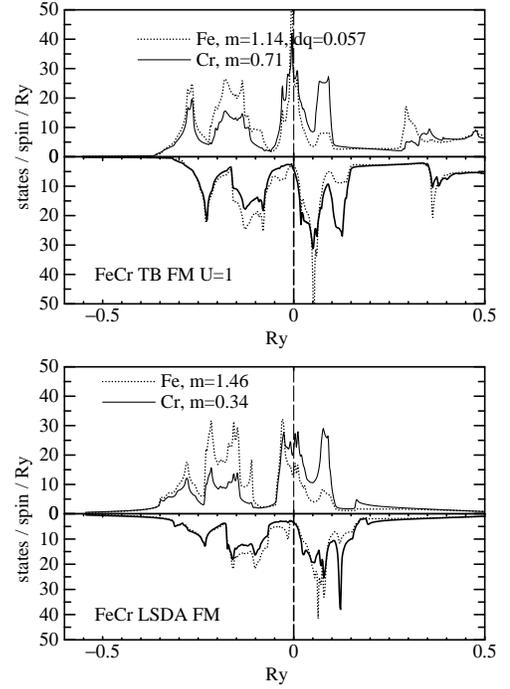}
\end{center}
\caption{Atom projected densities of states in ferromagnetic
FeCr. The self consistent tight-binding model is in close agreement
with the LSDA.}
\label{FM-FeCr-dos}
\end{figure}

\begin{figure}
\begin{center}
\includegraphics[scale=0.4,angle=0, trim=0 0 0 0]{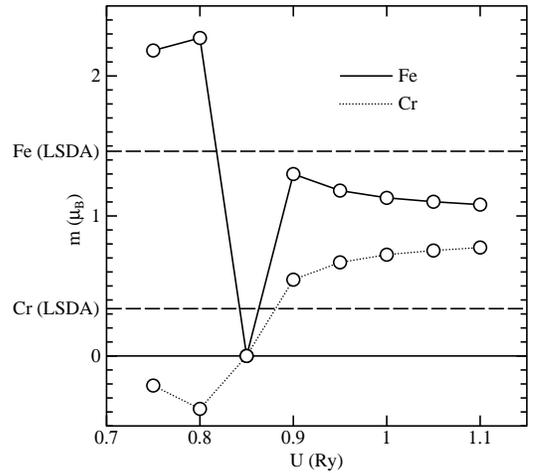}
\end{center}
\caption{Local magnetic moments (in $\muB$) on the Fe and Cr
atoms of B2 FeCr in the self consistent tight-binding model as a
function of the chosen Hubbard~$U$ parameter. For comparion the
estimated local moments from the LSDA are shown as horizontal
lines (of course, the Hubbard~$U$ is ``built-in'' to the LSDA and
cannot be varied. This may be regarded as an advantage and a
disadvantage: in the tight-binding one may observe the role of
parameters such as $I$ and $U$ by varying them). Note the
transition from antiferro- to ferromagnetism as $U$ is
increased. The limit of local charge neutrality leads to rather
smaller moments than found in the LSDA, but does predict the
correct magnetic ordering.}
\label{mom-vs-U}
\end{figure}

Figure~\ref{FM-FeCr-dos} shows the density of states in the self
consistent spin polarised tight-binding calculation employing a
Hubbard~$U$ of 1~Ry. The result is in close agreement with the
LSDA. The local moments on the Fe and Cr are 1.14~$\muB$ and
0.71~$\muB$, in reasonable accord with the estimated local
moments in the LSDA, namely 1.46~$\muB$ and
0.34~$\muB$. Figure~\ref{mom-vs-U} shows the local moments as a
function of the Hubbard~$U$ where we find an unphysical regime if
charge transfer is allowed to occur. In that case we find an
equal number of electrons in the spin up channel, while in the
spin down there is a larger population on the Fe than the Cr site
leading to antiferromagnetism.

\subsection{Chromium as a dilute impurity in iron}

\begin{figure}
\begin{center}
\includegraphics[scale=0.4,angle=0, trim=0 0 0 0]{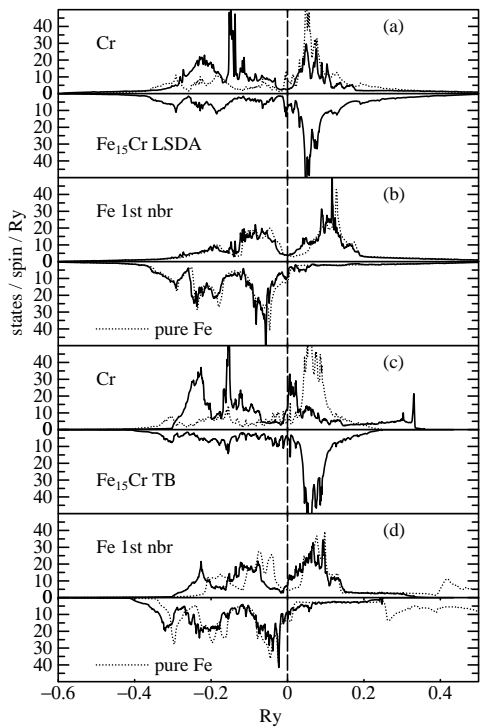}
\end{center}
\caption{Densities of states in an ordered Fe$_{15}$Cr alloy on
the \bcc\ lattice. (a) and (c) show the local density of states
projected onto the Cr~atom, respectively using LSDA and
tight-binding. The minority spin density of states in the lower
panel is repeated, using a dotted line, for comparison by
reflection about the abscissa. (b) and (d) show the local density
of states projected onto the Fe~atoms neighbouring the Cr, again
using LSDA and tight-binding respectively. A dotted line shows
the density of states in bulk Fe.}
\label{Fe15plusBulk-dos}
\end{figure}

\begin{figure}
\begin{center}
\includegraphics[scale=0.4,angle=0, trim=0 0 0 0]{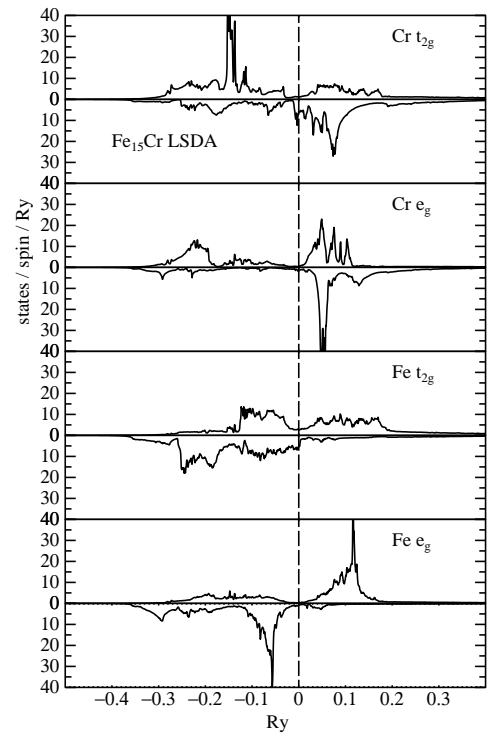}
\end{center}
\caption{LSDA local densities of states in Fe$_{15}$Cr projected into
the $t_{2g}$ and $e_g$ manifolds. This shows the non bonding
resonance on the Cr~impurity to originate from the $xy$, $yz$,
$zx$ and $d$-orbitals.}
\label{Fe15Cr-orbitals-LSDA-dos}
\end{figure}

\begin{figure}
\begin{center}
\includegraphics[scale=0.4,angle=0, trim=0 0 0 0]{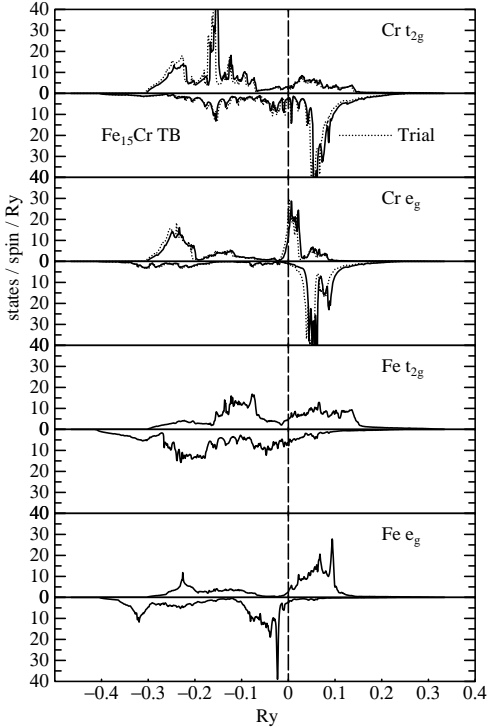}
\end{center}
\caption{As figure~\ref{Fe15Cr-orbitals-LSDA-dos}, but using the
tight-binding model. In the upper panel a dotted line shows the
density of states arising from a trial {\it input} non self consistent
consistent density with imposed moments on the Fe and Cr atoms
(see the text). Note, firstly, the excellent detailed agreement
with the LSDA in figure~\ref{Fe15Cr-orbitals-LSDA-dos} and
secondly the close similarity between the self consistent and non
self consistent densities of states.}
\label{Fe15Cr-orbitals-TB-dos}
\end{figure}

Whilst FeCr is ferromagnetic, a Cr atom in dilute concentration
in Fe becomes antiferromagnetically ordered with respect to the
Fe host atoms.\cite{Klaver06} We find that the self consistent
tight-binding model reproduces the LSDA remarkably well in
detail, and furthermore offers an explanation rather more readily
than the LSDA. We illustrate this using a unit cell of 16~sites
in the \bcc~Fe lattice, in one site of which an Fe~atom is
replaced with a Cr~atom. In figure~\ref{Fe15plusBulk-dos} we show
local densities of states projected onto the Cr and its
neighbouring Fe atoms, both using LSDA and tight-binding. Note
how the local density of states projected onto the Fe~atoms
neighbouring the Cr~impurity is hardly different from that of
bulk~Fe. It is curious that the Fe does not accommodate itself to
the presence of the Cr~impurity. On the other hand, the
Cr~projected density of states is greatly perturbed from its
bulk, as may be seen by comparison with
figure~\ref{AFM-Cr-dos}. The most prominent feature is a narrow
resonance in the occupied majority spins, which is almost
completely unhybridised with the neighbouring Fe~minority spins.
We show in figures~\ref{Fe15Cr-orbitals-LSDA-dos}
and~\ref{Fe15Cr-orbitals-TB-dos} the densities of states from
figure~\ref{Fe15plusBulk-dos} projected into the $t_{2g}$ and
$e_g$ manifolds. It becomes clear that this prominent feature
arises from strongly localised states of $xy$, $yz$ and $zx$
character. 

\section{The magnetic rigid band model}
\label{sec_RB-applications}

\begin{figure}
\begin{center}
\includegraphics[scale=0.4,angle=0, trim=0 0 0 0]{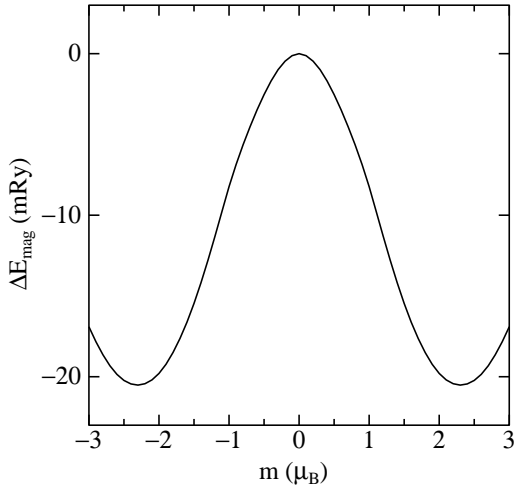}
\end{center}
\caption{Magnetic energy versus trial moment in the rigid band
Stoner model for \bcc-Fe (compare with figure~\ref{Stoner-LSDA}).
Curves of this type were first computed by Slater.\cite{Slater36b}}
\label{Fe-Stoner}
\end{figure}

It is clear from a comparison of figures~\ref{Fe15plusBulk-dos}
and~\ref{AFM-Cr-dos} that a rigid band approximation would be a
very poor description of alloying in the Fe--Cr system. A recent
calculation using the coherent potential approximation in the
LSDA has been made,\cite{Olsson06} but in this case, the
densities of states do not very well resemble those shown here in
figure~\ref{Fe15plusBulk-dos}. However, we may use the rigid band
model described in section~\ref{sec_RB-Stoner} in which the input
density is constructed having a trial moment. Indeed as seen in
figure~\ref{Fe15Cr-orbitals-TB-dos}, such a trial density (to be
described in detail below) gives a very faithful reproduction of
the self consistent density of states. In the simplest example,
that of the non magnetic density of states of Fe shown in
figure~\ref{NM-Fe-dos}, a plot of $\Delta\Emag$ from
equation~(\ref{emag}) versus $m$ is shown in
figure~\ref{Fe-Stoner}, having the characteristic
double-well\cite{Dudarev05} structure with minima at $m=2.3\muB$
and a magnetic energy of 21~mRy; these values may be compared
with those from the {\it self consistent} tight-binding
calculation in table~\ref{Fe_data}, {\it viz.}  $2.18\muB$ and
17~mRy. The small discrepancies arise from the self consistent
calculation allowing the shape of the spin densities of states to
be different from the input, non self consistent densities. As
mentioned in the the caption to figure~\ref{FM-Fe-dos}, above,
this is entirely due to the use of a non orthogonal tight-binding
model. 

\begin{figure}
\begin{center}
\includegraphics[scale=0.4,angle=0, trim=0 0 0 0]{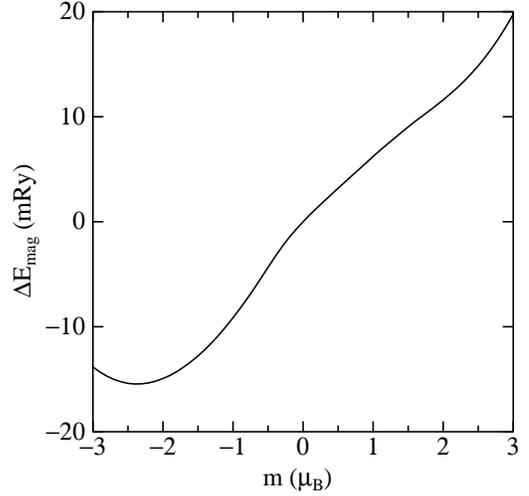}
\end{center}
\caption{Magnetic energy versus trial local Cr~impurity moment in
Fe$_{15}$Cr. Note there is only one, antiferromagnetic solution.}
\label{Fe15Cr-Stoner}
\end{figure}

We can now use this simple construction to interpret the
stability of the antiferromagnetic alignment of the Cr~impurity
in Fe. A trial spin polarised density is constructed by imposing
a moment of $+2.2\muB$ on each of the Fe~atoms and a trial moment
$m$ on the Cr~impurity. The associated bandstructure energy
difference is found to which $\frac{1}{4}I m^2$ is added, in
which we take $I=50$~mRy from table~\ref{tbl_parms}. The magnetic
energy plotted against $m$ is shown in
figure~\ref{Fe15Cr-Stoner}.

\begin{figure}
\begin{center}
\includegraphics[scale=0.4,angle=0, trim=0 0 0 0]{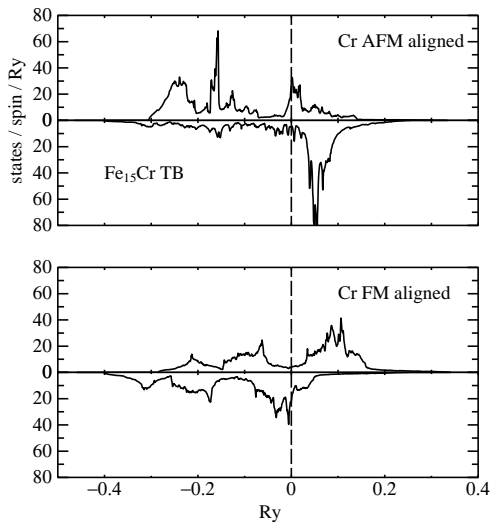}
\end{center}
\caption{Trial densities of states constructed by imposing
fixed magnetic moments onto the Cr~impurity in Fe$_{15}$Cr. As
already seen in figure~\ref{Fe15Cr-orbitals-TB-dos}, the trial
density for the antiferromagnetic alignment is very close to the
corresponding self consistent density of states. Because there is
no energy minimum at ferromagnetic alignment (see
figure~\ref{Fe15Cr-Stoner}) such a density cannot be achieved in
a self consistent procedure. This illustrates the usefulness of
this construction.}
\label{Cr-AFM-FM}
\end{figure}

Only one, antiferromagnetic, solution is found, having a local Cr
moment of 2.42$\muB$, which is close to the moment of 2.37$\muB$
found in the self consistent tight-binding calculation. (The
estimated Cr~local moment from our LSDA calculation is
2.08$\muB$.) Although there is no ferromagnetic solution, it is
instructive to plot the trial densities of states for trial local
moments of 2.37$\muB$ in both antiferro- and ferromagnetic
alignments. These are shown in figure~\ref{Cr-AFM-FM}. Neither
looks at all like the density of states of pure Cr in
figure~\ref{AFM-Cr-dos}; this is because to develop an
antiferromagnetic state requires the cooperation of two sub
lattices, which cannot be achieved by isolated Cr~atoms or small
clusters of these (say, fewer than nine atoms) in a \bcc-Fe
host. This is why it is the Cr density of states that has to
accommodate itself to the underlying Fe electronic structure, and
this lies at the heart of understanding the enthalpy of mixing
and the phase diagram in the Fe--Cr
system. Figure~\ref{Cr-AFM-FM} helps to explain why isolated
Cr~impurities do not align themselves ferromagnetically with the
host Fe. To do so would require a density of states essentially
that of pure {\it ferromagnetic} Cr, and this phase is unstable
with respect to the observed antiferromagnetic phase in Cr. The
alternative is to align antiferromagnetically, and this causes
the density of states to adopt a shape quite unlike that in pure
Cr while at the same time there is a complete lack of cooperation
from the very stable \bcc-Fe density of states, practically the
same as pure~Fe even on the Fe~atoms neighbouring the impurity.

\section{Origin of the repulsion between chromium impurites and the
enthalpy anomaly}
\label{FeCr-Enthalpy}

\begin{figure}
\begin{center}
\includegraphics[scale=0.4,angle=0, trim=0 0 0 0]{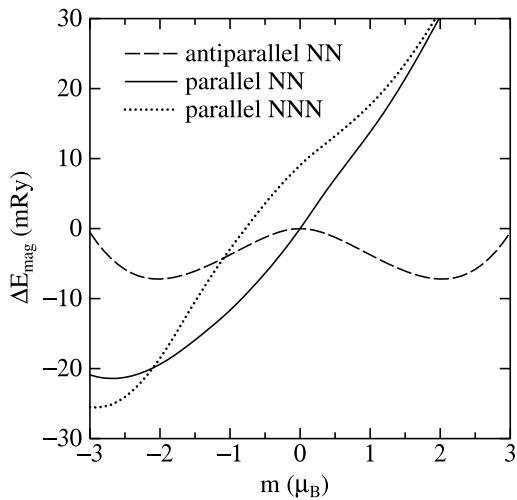}
\end{center}
\caption{Magnetic energy versus trial local Cr~impurity moment in
Fe$_{14}$Cr$_{2}$, having the two Cr~atoms as nearest neighbours,
NN, and next nearest neighbours, NNN. The solid line denotes the
energy of the pair of Cr~atoms having their spins aligned
parallel to each other. The broken line refers to the two Cr~spins being
aligned antiparallel to each other. The dotted line is the energy
in the case that the two Cr atoms are separated as next nearest
neighbours. The zero of energy in this plot is the energy of the
non magnetic NN case. Hence the graph shows the lowering of
energy as a result of moving the Cr atoms from NN to NNN
positions {\it as long as spin polarisation is permitted}. The
repulsion between Cr impurities is thereby revealed firstly as a
bandstructure effect and secondly as an effect of the magnetism.}
\label{Fe14Cr2-Stoner}
\end{figure}

Now we ask what is the stable magnetic structure of two
Cr~impurities placed as nearest neighbours in Fe? We go straight
to the predictions of the tight-binding model shown in
figure~\ref{Fe14Cr2-Stoner}. We make trial spin densities having
the Cr spins parallel or antiparallel to each other and plot the
magnetic energy as a function of their moment. In the case that
they are antiparallel and assuming the two moments to have same
magnitude, we find a double well as expected. The more stable
structure is for both spins to be aligned parallel to each other,
but to be antiferromagnetically aligned with the spins of the
Fe~host. In fact the antiparallel state is unstable and we find
that if the constraint is removed in a self consistent
calculation this reverts to the parallel state.

If the Cr atoms are placed at second neighbour positions, with
their spins aligned parallel to each other, we find an energy
versus magnetic moment very similar to that of the single
impurity in figure~\ref{Fe15Cr-Stoner}. In fact our LSDA and
tight-binding calculations (not presented here) show the
densities of states and magnetic moments to be very similar in
these two cases; indeed the LSDA local Cr moment is a little
larger in Fe$_{14}$Cr$_{2}$ than in Fe$_{15}$Cr as seen also in
the tight-binding model by comparing figures~\ref{Fe15Cr-Stoner}
and~\ref{Fe14Cr2-Stoner}. This latter figure now illustrates
rather clearly the origin of the repulsion between Cr impurites
in \bcc-Fe. The energy is {\it lower} when the atoms are placed
at next nearest neighbour positions as long as spin polarisation
is allowed; otherwise the energy ordering is reversed as is also
found using LSDA calculations.\cite{Klaver06} Furthermore since
the $\frac{1}{4}Im^2$ term is the same in both cases this is
clearly a {\it bandstructure effect}.

We can now offer a more detailed explanation for the anomaly in
the enthalpy of mixing of Cr in Fe. In most of the concentration
range Cr prefers to cluster together to allow sufficient atoms to
cooperate towards providing the two sublattices required to
establish the antiferromagnetic state. Hence the enthalpy of
mixing is positive and spinodal decomposition is
observed.\cite{Hyde95} Conversely at low concentrations, the Cr
may appear as isolated impurities stabilised by the change in
spin polarised density of states which has quite a large weight
at the bottom of the band as seen in
figure~\ref{Cr-AFM-FM}. These isolated impurites {\it repel} each
other, as already found by Klaver {\it et al.},\cite{Klaver06}
shown clearly in our figure~\ref{Fe14Cr2-Stoner}, hence at low
concentrations the enthalpy of mixing is negative, but only while
the concentration of Cr is sufficiently low for the Cr--Cr
repulsion to dominate. Our present modelling explains the nearest
neighbour repulsion in detail. The LSDA
calculations\cite{Klaver06} also showed that the Cr--Cr repulsion
extends to second neighbours and beyond, these longer ranged
interactions contribute significantly to the total repulsive
energy of a pair; furthermore they are present even when the
system is forced to be non spin polarised, when the nearest
neighbour repulsion collapses.  An explanation of the longer
ranged repulsion remains to be found in the bandstructure.

\section{Conclusions}
\label{conclusions}

We have described how itinerant magnetism can be incorporated
into our self consistent polarisable ion tight-binding
model. This results in an additional parameter, the Stoner $I$,
which we identify as minus twice the curvature of the exchange
and correlation energy as a function of magnetic moment. A first
order expansion of the Hohenberg--Kohn functional leads to the
rigid band Stoner--Slater model. We show that a very simple
parameterisation of the tight-binding model is possible that
gives a faithful reproduction of the energetics and electronic
structure of the LSDA. The parameters of the model are easily
transferable between the first row transiton metals and their
alloys. The simplest form of pair potential is quite adequate,
except in the case of Fe, to reproduce structural stability and
bulk modulus. Armed with this model we address outstanding
questions related to solution and clustering of Cr impurities in
\bcc-Fe. The rigid band, fixed moment approach proves to be very
useful in reproducing LSDA results and predicting magnetic
structure and energy of complex transition metal alloy
systems. This provides a powerful framework within which to
explore complex magnetic structures in transition metals
generally. The model is based in the correct physical picture,
namely itinerant magnetism resulting from a competition between
kinetic, or band, energy described by inter-site one electron
hopping matrix elements of the non self consistent tight-binding
Hamiltonian; and on-site exchange and correlation parameterised
through a single Stoner parameter.  Because the tight-binding
approximation is particularly simple and transparent we believe
that this approach will find a number of applications in this
area in the future.

\section*{Acknowledgement}

This work was supported by EPSRC under grants GR/S80165/01 and
GR/S81179/01.

\appendix
\section{Connection to \LDAU}
\label{App_A}

We may arrive at an expression similar to~(\ref{Eq_E2}) from the
starting point of the theory of \LDAU.\cite{Anisimov97} The usual
notation is to write $n^{\s}_m$ for the number of electrons or
occupation number in, say, a $d$-band with quantum number $m$
(not to be confused with the magnetic moment) and
spin $\s$. Then defining $U$ and $J$ as spheridised, orbital
independent Coulomb and exchange integrals, the on-site
electron--electron interaction energy is
\cite{Anisimov93,Anisimov97,Dudarev98,Petukhov03,Footnote5}
\begin{equation}
 \begin{split}
\EU&=\frac{1}{2}\,U\sum_{mm'\s}n^{\s}_m\,n^{-\s}_{m'}
   +\frac{1}{2}\left(U-J\right)\sum_{{mm'\s}\atop{m\ne m'}}
    n^{\s}_m\,n^{\s}_{m'}\\
   &=\frac{1}{2}U\ro^2-\frac{1}{2}J\sum_{\s}\left(\ro^{\s}\right)^2
     -\frac{1}{2}(U-J)\sum_{m\s}\left(n^{\s}_m\right)^2.\\
  \end{split}
\label{def_EU}
\end{equation}
The first line shows in its first term unlike spins interacting
through the Hubbard $U$, and in the second term like spin electrons
interacting through a Hubbard term reduced by an amount $J$
as explained at the end of section~\ref{sec_SCTB}. This term
explicitly requires $m\ne m'$ in the sum: as two electrons cannot
occupy the same state according to the Pauli principle this would
otherwise give an interaction between an electron and
itself. Hence the {\it on-site} electron--electron interaction
properly includes the so called self interaction correction
present in Hartree--Fock theory but not in the
LSDA.\cite{Footnote2}
The second line\cite{Petukhov03} follows directly after some
algebra, expressing
$$\ro=\sum_{m\s}n^{\s}_m,\hskip 12pt \ro^{\s}=\sum_{m}n^{\s}_m.$$ The
three terms resulting in the second line of~(\ref{def_EU}) are
respectively a direct Coulomb term, an exchange term
and a term which is of lower order of magnitude compared to the
first two and which would amount to admitting an {\it orbital
dependent} potential. In the spirit of the LSDA we
neglect\cite{Footnote3} this last term and by differentiation we
find for the potential seen by an electron with spin $\s$ as a
result of electron--electron interaction,
$$V^{\s} = \frac{\partial\EU}{\partial \ro^{\s}} = U\ro - J\ro^{\s}$$ 
and so the exchange splitting between up
and down spin energy levels is approximately
$\Delta\varepsilon \sim
\Vup-\Vdn=-J\left(\rup-\rdn\right)=-Jm.$ After some
further algebra again neglecting the third term
in~(\ref{def_EU}) we may also write
\begin{equation}
\EU=\frac{1}{2}\left(U-\frac{1}{2}J\right)\ro^2-\frac{1}{4}Jm^2
\label{Eq_EU}
\end{equation}
which is equivalent to our expression~(\ref{Eq_E2}) for $\Etwo$
in section~\ref{sec_SCTB} after identifying the exchange integral
$J$ with the Stoner parameter $I$. Note, however, that $\EU$ is
not an energy to second order in any charge density difference,
but it could be cast into such a form if we make an expansion of
the total energy in a generalised mean field multiband Hubbard
model. We wish to emphasise two points here. ({\it i\/}) Both
exchange {\it and} correlation are contained in
equations~(\ref{Eq_E2}) and~(\ref{Eq_EU}), the effective Coulomb
integral being reduced to $U-\frac{1}{2}J$ by exchange. Indeed it
is well known that the exchange-only Kohn--Sham--Gaspar potential
gives a poor description of itinerant magnetism by overestimating
the tendency to magnetism in transition
metals.\cite{Gunnarsson76} ({\it ii\/}) As in LSDA,
equations~(\ref{Eq_E2}) and~(\ref{Eq_EU}) are functionals of the
spin density only and lead to orbital independent potentials. It
is clear, though, from the foregoing how to recover the self
interaction correction (at least in on-site terms in the
Hamiltonian) in a tight-binding context in which the potential
seen by an electron is orbital dependent.
\nocite{Sanna07}
        
\section{Non orthogonal self consistent tight-binding}
\label{App_B}

There is a number of benefits of adopting a non orthogonal
tight-binding basis. It is widely believed to result in a more
transferable model. In addition it admits the concept of {\it
bond charge.}\cite{Pettifor95} As we now demonstrate this allows
the self consistency to adjust the {\it hopping integrals} as
well as on-site matrix elements of the Hamiltonian. We recall that
our self consistent polarisable ion tight-binding
model\cite{FinnisMRS98,Finnis98,Fabris00} is couched in terms of
multipole moments of charge with respect to neutral, spherical
atoms having $q_{\R}^0$ valence electrons. Hence the self
consistent charge transfer to a site labelled by its position
$\R$ in units of electron charge, $e$, is
\begin{equation}
\delta q_{\R}=q_{\R}-q_{\R}^0\equiv\Qz.
\label{def_q}
\end{equation}
Higher moments of the charge develop as a
result of crystal field splitting and these are denoted $\Q$, in
which $L$ is a composite index subsuming both angular momenta:
$\L=\{\ell m\}$. The Madelung potential (energy) at site $\R$ due to
multipoles at sites $\Rp$ is
\begin{equation}
V^M_{\R\L}=e^2\sum_{\Rp\ne\R}\sum_{\Lp}B_{\L\Lp}
             \left(\Rp-\R\right)\Qp.
\label{def_V}
\end{equation}
${\bf B}$ is a generalised Madelung
matrix,\cite{FinnisMRS98,Finnis03} related to the structure
constants of LMTO theory.\cite{Andersen85} For monopole
interactions, we write
$$B_{00}\left(\Rp-\R\right)=\frac{1}{\left\vert\Rp-\R\right\vert}\equiv
U_{\R\Rp}.$$ The transfer of charge is resisted by a ``Hubbard
potential,''
$$V^U_{\R}=U_{\R}\Qz.$$ 
In the {\it orthogonal} self consistent tight-binding model, these
potentials are used to adjust the {\it on-site} matrix elements
of the Hamiltonian, both on-site energies and off-diagonal
crystal field terms. The increments to the Hamiltonian are
$$V_{\R\L\R\Lp}=V^U_{\R}\delta_{\L\Lp}+\sum_{\Lpp}V^M_{\R\Lpp}\M\gaunt$$ in
which $\gaunt$ are the Gaunt coefficents that enforce the
selection rules and $\M$ are new parameters controlling the
strength of the crystal field
splitting.\cite{FinnisMRS98,Finnis98,Finnis03} These may be
adjusted, for example, to reproduce crystal field splittings in
{\it ab initio} bandstructures or dipole moments in molecules.

If we include an overlap matrix $S_{\R\L\Rp\Lp}$, then solving
the generalised eigenproblem leads to normalised eigenvectors
$\C$ and the charge at site $\R$ is
\begin{eqnarray*}
q_{\R}&=&\frac{1}{2}\sum_{n\k}f_{n\k}\sum_{\Rp\Lp\L}
         \left(\Cb\Sk\C+{\hbox{c.c.}}\right)\\
      &=&\sum_{n\k}f_{n\k}\sum_{\Lp\L}\left|\C\right|^2\\
      &+&\frac{1}{2}\sum_{n\k}f_{n\k}\sum_{\Rp\Lp\L}
         \left(\Cb\Ok\C+{\hbox{c.c.}}\right).\\
\end{eqnarray*}
Here, a bar and ``c.c.'' imply complex conjugation. $f_{n\k}$ are
occupation numbers\cite{Footnote1}
of the state at wavevector $\k$ and band index
$n$, as used say in Fermi--Dirac or generalised Gaussian
Brillouin zone integration,\cite{Kresse96} or the linear tetrahedron
method.\cite{Jepsen71} The final term amounts to a {\it bond
charge} which is absent in orthogonal tight-binding models. To
extract the bond charge explicitly, we have defined ${\bf O}={\bf
S}-{\bf 1}$ and since the norm is conserved separately at each
$\k$-point, we work with Bloch transformed matrices, such that,
for example,
$$\Sk=\sum_{\bf T}S_{\R+{\bf T}\L\Rp\Lp}\,\e^{{\rm i}\k\cdot{\bf T}},$$
where ${\bf T}$ are the translation vectors of the crystal
lattice.

For simplicity we allow the overlap to make contributions only to
the monopole moments of the charge; higher moments are defined as
in the orthogonal case so that for $\ell>0$ we
have,\cite{FinnisMRS98,Finnis98,Fabris00,Finnis03}
$$\Q=\sum_{n\k}f_{n\k}\sum_{\Lp\Lpp}
     {\bar C}^{n\k}_{\R\Lp} C^{n\k}_{\R\Lpp}\M\gaunt.
$$

\begin{figure}
\begin{center}
\includegraphics[scale=0.5,angle=0, trim=0 0 0 0]{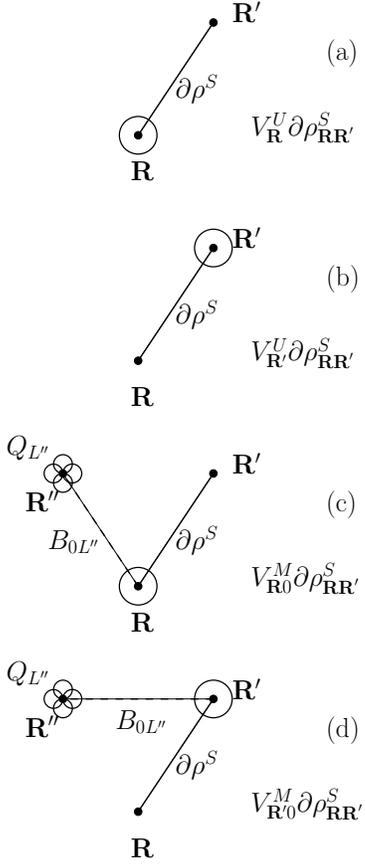}
\end{center}
\caption{Hubbard and Madelung contributions to the force on atom
$\R$. Circles are intended to represent changes in monopoles
arising from the displacement of the atom at $\R$, which modifies
the charge on both sites $\R$ and $\Rp$ through the scaling of
the overlap matrix elements with bond length.}
\label{Overlap}
\end{figure}

We now find increments to the hopping integrals as a result of
the self consistent redistribution of bond charge. These
are\cite{Finnis03}
$$V^{\k}_{\R\L\Rp\Lp}=\frac{1}{2}\left(D_{\R}+D_{\Rp}\right)\Ok$$
where
$$D_{\R}=V^U_{\R}+\sum_{\Rp}U_{\R\Rp}\Qzp.$$ 
To preserve the norm these need to be updated directly into the
Bloch transformed Hamiltonian.  Note that only monopole terms
enter here as a result of our definition of the higher multipoles
without reference to the overlap. $D_{\R}$ is the sum of Hubbard
and point-charge Madelung potentials at site $\R$.

There are also new terms in the interatomic forces. According to
the Hellmann--Feynman theorem the force is obtained from the
derivative of the energy, taken while keeping the wavefunction
frozen. In an orthogonal tight-binding model multipole moments do
not change under this constraint when the atom at $\R$ is
displaced; hence the only contribution to the force from self
consistent, second order terms in the energy is the classical
electrostatic term,
$$
 {\bf F}_{\R}^{\hbox{es}}=
  -\frac{1}{2}e^2\sum_{{\Rp\ne\Rpp}\atop{\Lp\Lpp}}
  \Qp\,\frac{\partial
  B_{\Lp\Lpp}\left(\Rpp-\Rp\right)}{\partial\R}
  \,\Qpp.
$$
However in a {\it non orthogonal} model, even at fixed
eigenvectors, displacement of an atom will lead to changes in
the bond charges with its neighbouring atoms as a result of the
changes in the overlap matrix elements. There are two new
contributions to the interatomic force. Since we are concerned
with derivates of the overlap matrix, we will require the
quantity
$$\partial\rho^S_{\R\Rp}=
  \sum_{\L\Lp}\partial\rho^S_{\R,\R\L\Rp\Lp}$$
where
\begin{eqnarray*}
\partial\rho^S_{\R,\R\L\Rp\Lp}
&=&-\partial\rho^S_{\Rp,\R\L\Rp\Lp}\\
=\frac{1}{2}\sum_{n\k}&f_{n\k}&
   \left(\Cb\frac{\partial\Sk}{\partial\R}\,\Cp+{\hbox{c.c.}}\right).\\
\end{eqnarray*}
Then for the first contribution we find
$${\bf F}^U_{\R}=-\sum_{\Rp}\left(V^U_{\R}+V^U_{\Rp}\right)
                 \partial\rho^S_{\R\Rp}
$$
and for the Madelung contribution,
$${\bf F}^M_{\R}=-\sum_{\Rp}\left(V^M_{\R 0}+V^M_{\Rp 0}\right)
                 \partial\rho^S_{\R\Rp}.
$$
$V^U_{\R}$ is the Hubbard potential, and $V^M_{\R 0}$ is the
$\ell=0$ component of the electrostatic potential~(\ref{def_V})
seen at $\R$.  These two contributions to the interatomic force
are open to a quite simple interpretation if we make reference to
figure~\ref{Overlap}.

When the atom at $\R$ moves, its own monopole moment changes by
virtue of overlap with an atom at $\Rp$. This leads to a change
in Hubbard potential (energy) at site $\R$ and hence a force
(figure~\ref{Overlap}(a)). This change in monopole moment at $\R$
will result in a modified electrostatic interaction with a
multipole moment at a third site $\Rpp$ (including the
possibility $\Rpp=\Rp$) described by the matrix element
$B_{0\Lpp}\left(\Rpp-\R\right)$. This leads to the first Madelung
contribution, shown in figure~\ref{Overlap}(c).  The same
movement also induces a change in the monopole moment at site
$\Rp$ giving rise to the second Hubbard contribution, shown in
figure~\ref{Overlap}(b). The second Madelung contribution,
illustrated in figure~\ref{Overlap}(d), corresponds to the force
associated with the electrostatic interaction between a multipole
at $\Rpp$ (admitting the possibility that $\Rpp=\R$) and the
modified charge at $\Rp$ through the Madelung matrix element
$B_{0\Lpp}\left(\Rpp-\Rp\right)$.

\def\JPCM{J.~Phys.:~Condens.~Matter.}
\def\PRB{Phys.~Rev.~B}
\def\PRL{Phys.~Rev.~Lett.}

\bibliography{Me,Footnote}

\begin{thebibliography}{68}
\bibitem[{Foo({\natexlab{a}})}]{Footnote4}
\bibinfo{note}{This is the same $I$ as defined by Janak\cite{Janak77} averaged
  inside the atomic sphere as long as we admit a misprint in his equation~(4):
  the ``two'' should become a ``half.''}.
\bibitem[{Foo({\natexlab{c}})}]{Footnote5}
\bibinfo{note}{Anisimov {\it et al.}\cite{Anisimov93} give a particularly clear
  derivation of this expression for $\EU$, but note a misprint in their
  equation~(8): the $E$ is not the same as the $E$ in their equation~(5) but is
  intended as a substitute for the {\it last term} in the latter equation
  taking into account exchange in addition to the direct Coulomb interaction.}
\bibitem[{Foo({\natexlab{d}})}]{Footnote2}
\bibinfo{note}{The more general form \begin{align*}
  \EU&=\frac{1}{2}\sum_{mm'\s}U_{mm'}\,n^{\s}_m\,n^{-\s}_{m'}\\
  &+\frac{1}{2}\sum_{mm'\s} \left(U_{mm'}-J_{mm'}\right)n^{\s}_m\,n^{\s}_{m'}
  \end{align*} accounts for self interaction implicitly since $U_{mm}=J_{mm}$.
  It is reasonable in tight-binding as in the atomic spheres approximation to
  adopt the spherical averaged $U$ and $J$ in which case $m'=m$ must be
  excluded explicitly in the sum over like spins.}
\bibitem[{Foo({\natexlab{f}})}]{Footnote1}
\bibinfo{note}{It is to be understood that these occupation numbers are divided
  by the total number of {$\bf k$}-points in the Brillouin zone, or scaled by
  the number and degeneracy of {$\bf k$}-points in the irreducible zone.}
\end{thebibliography}

\end{document}